\documentstyle[axodraw]{article}

\begin{document}

\begin{center}
{\large \bf Contact interactions in the four-fermion processes at LEP2 and
HERA} \\

\vspace{4mm}

       F.Berends \\
       CERN \\
       CH-1211 Geneva 23, Switzerland\\
\vspace{4mm}        
         M.Dubinin \\
       Institute of Nuclear Physics, Moscow State University \\
       119899 Moscow, Russia
\end{center}

\begin{abstract}

We consider the possibility of experimental observation of 
flavor-diagonal and helicity conserving contact terms in the
four fermion reactions $e^+ e^- \rightarrow e^+ e^- \mu^+ \mu^-$
and $e^- q \rightarrow e^- \mu^+ \mu^- q$ at LEP2 and HERA.

\end{abstract}

\section{Introduction}

Speculations about the composite nature of leptons and quarks have
been developed during a long period of time \cite{general}. With the
help of compositeness of fundamental fermions one could hope to
inderstand a number of principal features of the Standard Model
scheme such as the structure of fermion generations, mass spectrum
of fermions and the symmetry breaking scenario.

A large number of phenomenological studies of the possibility to
observe the sigatures of compositeness at the new generation of
$e^+ e^-$, $ep$ and $p \bar p$ colliders exist (see 
\cite{ee,ep,pp}
and references therein). One can imagine a simplified picture
when leptons and quarks consist of some pointlike particles (preons)
bound by some new interaction (metacolor force) which is probably confining 
and 
become strong at some energy scale $\Lambda$. If at the new colliders the
momentun transfer exceeds $\Lambda$, leptons and quarks would interact
in a manner completely different from their pointlike low energy 
structure, showing directly the hard 
scattering processes of the constituents. At the energies less
than $\Lambda$ one could observe some indications to the constituent
dynamics (residual effective interactions) and describe this regime 
in the framework of some effective lagrangian approach. This effective
lagrangian is given by a Standard Model lagrangian and some 
operators of higher dimension involving the fields of the SM. For
instance, the simplest effective term of this type is given by
dimension six four-fermion operator $(\psi \gamma \psi)(\psi \gamma
\psi)$ multiplied by $g^2/\Lambda^2$ giving the effective term
with correct dimension four. The strength of such nonrenormalisable
effective interactions is determined by a dimensionless coupling $g$
and powers of the compositeness scale $\Lambda$.

\section{Distributions in the Standard Model with $LL-$ contact term
for the process $e^+ e^- \rightarrow e^+ e^- \mu^+ \mu^-$} 

\subsection{Parametrization of contact interactions}

We are using helicity conserving contact interactions of the form 
\cite{Eichten}
\begin{equation}
L_c=\frac{g^2}{2\Lambda^2}(\eta_{LL} \bar \psi_L \gamma_{\mu} \psi_L
                                     \bar \psi_L \gamma^{\mu} \psi_L
                          +  \eta_{RR} \bar \psi_R \gamma_{\mu} \psi_R 
                                     \bar \psi_R \gamma^{\mu} \psi_R
                          + 2 \eta_{LR} \bar \psi_R \gamma_{\mu} \psi_R
                                     \bar \psi_L \gamma^{\mu} \psi_L)
\end{equation}
where $g^2/4\pi=1$, $|\eta|=1$ and $\psi_{L,R}=(1 \mp \gamma_5)\psi/2$.
In the case of positive $\eta$ the
first and second terms are denoted by $LL+$ and $RR+$, if $\eta$ is
negative they are denoted by $LL-$ and $RR-$ correspondingly \cite{ee}.
Particular choice of $\eta_i$ gives $VV$ and $AA$ (vector-vector and 
axial-axial) current interactions.
In the following we choose the $LL-$ contact term. No qualitative
difference in the results appears if we choose any other variant from the
six possible. Previous analyses performed in \cite{Eichten,Schrempp}
for the reactions $e^+ e^- \rightarrow e^+ e^-,\, e^+ e^- \rightarrow
\mu^+ \mu^-$ showed that the effect of $LL$ and $RR$ terms is typically 
several times smaller than the effect of $VV$, $AA$ terms.

\begin{figure}[t]
\begin{center}
{\def\chepscale{1.0} 
\unitlength=\chepscale pt
\SetWidth{0.7}      
\SetScale{\chepscale}
\tiny    
\begin{picture}(50,90)(0,0)
\Text(7.7,77.8)[r]{$e1$}
\ArrowLine(8.0,77.8)(30.7,77.8) 
\Text(42.3,77.8)[l]{$e1$}
\ArrowLine(30.7,77.8)(42.0,77.8) 
\Text(30.3,67.0)[r]{$A$}
\DashLine(30.7,77.8)(30.7,56.2){3.0} 
\Text(42.3,56.2)[l]{$E2$}
\ArrowLine(42.0,56.2)(30.7,56.2) 
\Text(29.2,45.4)[r]{$e2$}
\ArrowLine(30.7,56.2)(30.7,34.6) 
\Text(42.3,34.6)[l]{$e2$}
\ArrowLine(30.7,34.6)(42.0,34.6) 
\Text(30.3,23.8)[r]{$A$}
\DashLine(30.7,34.6)(30.7,13.0){3.0} 
\Text(7.7,13.0)[r]{$E1$}
\ArrowLine(30.7,13.0)(8.0,13.0) 
\Text(42.3,13.0)[l]{$E1$}
\ArrowLine(42.0,13.0)(30.7,13.0) 
\Text(25,0)[b] {diagr.1}
\end{picture} \ 
\begin{picture}(50,90)(0,0)
\Text(7.7,77.8)[r]{$e1$}
\ArrowLine(8.0,77.8)(30.7,77.8) 
\Text(42.3,77.8)[l]{$e1$}
\ArrowLine(30.7,77.8)(42.0,77.8) 
\Text(30.3,67.0)[r]{$A$}
\DashLine(30.7,77.8)(30.7,56.2){3.0} 
\Text(42.3,56.2)[l]{$e2$}
\ArrowLine(30.7,56.2)(42.0,56.2) 
\Text(29.2,45.4)[r]{$e2$}
\ArrowLine(30.7,34.6)(30.7,56.2) 
\Text(42.3,34.6)[l]{$E2$}
\ArrowLine(42.0,34.6)(30.7,34.6) 
\Text(30.3,23.8)[r]{$A$}
\DashLine(30.7,34.6)(30.7,13.0){3.0} 
\Text(7.7,13.0)[r]{$E1$}
\ArrowLine(30.7,13.0)(8.0,13.0) 
\Text(42.3,13.0)[l]{$E1$}
\ArrowLine(42.0,13.0)(30.7,13.0) 
\Text(25,0)[b] {diagr.2}
\end{picture} \ 
}
 \\
{\def\chepscale{1.0} 
\unitlength=\chepscale pt
\SetWidth{0.7}      
\SetScale{\chepscale}
\tiny    
\begin{picture}(50,90)(0,0)
\Text(7.7,77.8)[r]{$e1$}
\ArrowLine(8.0,77.8)(30.7,77.8) 
\Text(42.3,77.8)[l]{$e1$}
\ArrowLine(30.7,77.8)(42.0,77.8) 
\Text(30.3,67.0)[r]{$A$}
\DashLine(30.7,77.8)(30.7,56.2){3.0} 
\Text(42.3,56.2)[l]{$E2$}
\ArrowLine(42.0,56.2)(30.7,56.2) 
\Text(29.2,45.4)[r]{$e2$}
\ArrowLine(30.7,56.2)(30.7,34.6) 
\Text(42.3,34.6)[l]{$e2$}
\ArrowLine(30.7,34.6)(42.0,34.6) 
\Text(30.3,23.8)[r]{$X$}
\DashLine(30.7,34.6)(30.7,13.0){3.0} 
\Text(7.7,13.0)[r]{$E1$}
\ArrowLine(30.7,13.0)(8.0,13.0) 
\Text(42.3,13.0)[l]{$E1$}
\ArrowLine(42.0,13.0)(30.7,13.0) 
\Text(25,0)[b] {diagr.1}
\end{picture} \ 
\begin{picture}(50,90)(0,0)
\Text(7.7,77.8)[r]{$e1$}
\ArrowLine(8.0,77.8)(30.7,77.8) 
\Text(42.3,77.8)[l]{$e1$}
\ArrowLine(30.7,77.8)(42.0,77.8) 
\Text(30.3,67.0)[r]{$A$}
\DashLine(30.7,77.8)(30.7,56.2){3.0} 
\Text(42.3,56.2)[l]{$e2$}
\ArrowLine(30.7,56.2)(42.0,56.2) 
\Text(29.2,45.4)[r]{$e2$}
\ArrowLine(30.7,34.6)(30.7,56.2) 
\Text(42.3,34.6)[l]{$E2$}
\ArrowLine(42.0,34.6)(30.7,34.6) 
\Text(30.3,23.8)[r]{$X$}
\DashLine(30.7,34.6)(30.7,13.0){3.0} 
\Text(7.7,13.0)[r]{$E1$}
\ArrowLine(30.7,13.0)(8.0,13.0) 
\Text(42.3,13.0)[l]{$E1$}
\ArrowLine(42.0,13.0)(30.7,13.0) 
\Text(25,0)[b] {diagr.2}
\end{picture} \ 
\begin{picture}(50,90)(0,0)
\Text(7.7,77.8)[r]{$e1$}
\ArrowLine(8.0,77.8)(30.7,77.8) 
\Text(42.3,77.8)[l]{$e1$}
\ArrowLine(30.7,77.8)(42.0,77.8) 
\Text(30.3,67.0)[r]{$X$}
\DashLine(30.7,77.8)(30.7,56.2){3.0} 
\Text(42.3,56.2)[l]{$E2$}
\ArrowLine(42.0,56.2)(30.7,56.2) 
\Text(29.2,45.4)[r]{$e2$}
\ArrowLine(30.7,56.2)(30.7,34.6) 
\Text(42.3,34.6)[l]{$e2$}
\ArrowLine(30.7,34.6)(42.0,34.6) 
\Text(30.3,23.8)[r]{$A$}
\DashLine(30.7,34.6)(30.7,13.0){3.0} 
\Text(7.7,13.0)[r]{$E1$}
\ArrowLine(30.7,13.0)(8.0,13.0) 
\Text(42.3,13.0)[l]{$E1$}
\ArrowLine(42.0,13.0)(30.7,13.0) 
\Text(25,0)[b] {diagr.3}
\end{picture} \ 
\begin{picture}(50,90)(0,0)
\Text(7.7,77.8)[r]{$e1$}
\ArrowLine(8.0,77.8)(30.7,77.8) 
\Text(42.3,77.8)[l]{$e1$}
\ArrowLine(30.7,77.8)(42.0,77.8) 
\Text(30.3,67.0)[r]{$X$}
\DashLine(30.7,77.8)(30.7,56.2){3.0} 
\Text(42.3,56.2)[l]{$e2$}
\ArrowLine(30.7,56.2)(42.0,56.2) 
\Text(29.2,45.4)[r]{$e2$}
\ArrowLine(30.7,34.6)(30.7,56.2) 
\Text(42.3,34.6)[l]{$E2$}
\ArrowLine(42.0,34.6)(30.7,34.6) 
\Text(30.3,23.8)[r]{$A$}
\DashLine(30.7,34.6)(30.7,13.0){3.0} 
\Text(7.7,13.0)[r]{$E1$}
\ArrowLine(30.7,13.0)(8.0,13.0) 
\Text(42.3,13.0)[l]{$E1$}
\ArrowLine(42.0,13.0)(30.7,13.0) 
\Text(25,0)[b] {diagr.4}
\end{picture} \ 
}
 \\
{\def\chepscale{1.0} 
\unitlength=\chepscale pt
\SetWidth{0.7}      
\SetScale{\chepscale}
\tiny    
\begin{picture}(50,90)(0,0)
\Text(7.7,67.0)[r]{$e1$}
\ArrowLine(8.0,67.0)(19.3,67.0) 
\Text(24.8,69.8)[b]{$e1$}
\ArrowLine(19.3,67.0)(30.7,67.0) 
\Text(42.3,77.8)[l]{$e1$}
\ArrowLine(30.7,67.0)(42.0,77.8) 
\Text(30.3,56.2)[r]{$X$}
\DashLine(30.7,67.0)(30.7,45.4){3.0} 
\Text(42.3,56.2)[l]{$e2$}
\ArrowLine(30.7,45.4)(42.0,56.2) 
\Text(42.3,34.6)[l]{$E2$}
\ArrowLine(42.0,34.6)(30.7,45.4) 
\Text(19.0,45.4)[r]{$A$}
\DashLine(19.3,67.0)(19.3,23.8){3.0} 
\Text(7.7,23.8)[r]{$E1$}
\ArrowLine(19.3,23.8)(8.0,23.8) 
\Line(19.3,23.8)(30.7,23.8) 
\Text(42.3,13.0)[l]{$E1$}
\ArrowLine(42.0,13.0)(30.7,23.8) 
\Text(25,0)[b] {diagr.1}
\end{picture} \ 
\begin{picture}(50,90)(0,0)
\Text(7.7,67.0)[r]{$e1$}
\ArrowLine(8.0,67.0)(19.3,67.0) 
\Line(19.3,67.0)(30.7,67.0) 
\Text(42.3,77.8)[l]{$e1$}
\ArrowLine(30.7,67.0)(42.0,77.8) 
\Text(19.0,56.2)[r]{$A$}
\DashLine(19.3,67.0)(19.3,45.4){3.0} 
\Line(19.3,45.4)(30.7,45.4) 
\Text(42.3,56.2)[l]{$E1$}
\ArrowLine(42.0,56.2)(30.7,45.4) 
\Text(17.9,34.6)[r]{$e1$}
\ArrowLine(19.3,45.4)(19.3,23.8) 
\Text(7.7,23.8)[r]{$E1$}
\ArrowLine(19.3,23.8)(8.0,23.8) 
\Text(24.8,24.5)[b]{$X$}
\DashLine(19.3,23.8)(30.7,23.8){3.0} 
\Text(42.3,34.6)[l]{$e2$}
\ArrowLine(30.7,23.8)(42.0,34.6) 
\Text(42.3,13.0)[l]{$E2$}
\ArrowLine(42.0,13.0)(30.7,23.8) 
\Text(25,0)[b] {diagr.2}
\end{picture} \ 
\begin{picture}(50,90)(0,0)
\Text(7.7,67.0)[r]{$e1$}
\ArrowLine(8.0,67.0)(19.3,67.0) 
\Line(19.3,67.0)(30.7,67.0) 
\Text(42.3,77.8)[l]{$e1$}
\ArrowLine(30.7,67.0)(42.0,77.8) 
\Text(19.0,56.2)[r]{$A$}
\DashLine(19.3,67.0)(19.3,45.4){3.0} 
\Text(7.7,45.4)[r]{$E1$}
\ArrowLine(19.3,45.4)(8.0,45.4) 
\Text(24.8,48.2)[b]{$e1$}
\ArrowLine(30.7,45.4)(19.3,45.4) 
\Text(42.3,56.2)[l]{$E1$}
\ArrowLine(42.0,56.2)(30.7,45.4) 
\Text(30.3,34.6)[r]{$X$}
\DashLine(30.7,45.4)(30.7,23.8){3.0} 
\Text(42.3,34.6)[l]{$e2$}
\ArrowLine(30.7,23.8)(42.0,34.6) 
\Text(42.3,13.0)[l]{$E2$}
\ArrowLine(42.0,13.0)(30.7,23.8) 
\Text(25,0)[b] {diagr.3}
\end{picture} \ 
\begin{picture}(50,90)(0,0)
\Text(7.7,67.0)[r]{$e1$}
\ArrowLine(8.0,67.0)(19.3,67.0) 
\Text(24.8,67.7)[b]{$X$}
\DashLine(19.3,67.0)(30.7,67.0){3.0} 
\Text(42.3,77.8)[l]{$e2$}
\ArrowLine(30.7,67.0)(42.0,77.8) 
\Text(42.3,56.2)[l]{$E2$}
\ArrowLine(42.0,56.2)(30.7,67.0) 
\Text(17.9,56.2)[r]{$e1$}
\ArrowLine(19.3,67.0)(19.3,45.4) 
\Line(19.3,45.4)(30.7,45.4) 
\Text(42.3,34.6)[l]{$e1$}
\ArrowLine(30.7,45.4)(42.0,34.6) 
\Text(19.0,34.6)[r]{$A$}
\DashLine(19.3,45.4)(19.3,23.8){3.0} 
\Text(7.7,23.8)[r]{$E1$}
\ArrowLine(19.3,23.8)(8.0,23.8) 
\Line(19.3,23.8)(30.7,23.8) 
\Text(42.3,13.0)[l]{$E1$}
\ArrowLine(42.0,13.0)(30.7,23.8) 
\Text(25,0)[b] {diagr.4}
\end{picture} \ 
}
\end{center}
\caption{Subset of 10 t-channel diagrams for the process
          $e^+ e^- \rightarrow e^+ e^- \mu^+ \mu^-$. SM diagrams are in
the first row, in the second and third rows the X-particle exchange
corresponds to some contact interaction.}
\end{figure}
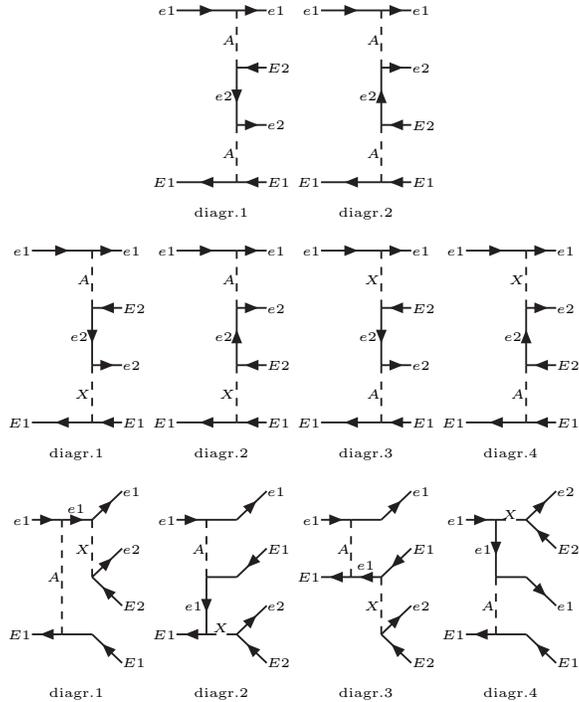

\subsection{Search strategies and kinematical cuts}

Careful analysis is necessary in the Standard Model  
$W$ and Higgs boson production for the definition of the signal versus
the background in the four fermion final state \cite{4f}.
Usually it is more difficult to separate the small signal of new physics,
strongly restricted by the data from independent experiments, in the exclusive
multiparticle final state. 
One can propose two contact terms search strategies. In the framework of
the first strategy we impose loose kinematical cuts on the four fermion
final state, the number of identifiable events is large enough, the 
contribution
of the contact term in addition to the SM distribution is small, but the
statistical error is also small and one can hope to observe a deviation
from the SM cross section in the high statistics experiment. Especially 
interesting is the case if the interference of SM and $LL-$ defined 
amplitudes is large. In the framework
of the second strategy we impose stringent kinematical cuts, the number
of events is very small, but the contribution of the contact term in addition
to the SM distribution can be large and
one can hope to observe large deviation in the experiment with
a small number of events. Generally speaking it is difficult to say in
advance what strategy would be better.

 For the first and second strategy we are using the following 
cuts: 

\noindent {\bf Set I (Loose cuts):}\\
muon pair mass cut $M(\mu^+ \mu^-) \geq$ 30, 60, 85 GeV (three cases)\\
final muon energy cut $E \geq$ 10 GeV\\
final muon angle with the beams $\vartheta \geq$ 10 degrees 

\noindent {\bf Set II (Strong cuts):}\\
muon pair mass cut $M(\mu^+ \mu^-) \geq$ 30, 60, 85 GeV (three cases)\\
electron pair mass cut $M(e^+ e^-) \geq$ 3.16 GeV \\
electrons angular cut with the beam $\vartheta \geq$ 10 degrees \\
final lepton energy cut $E \geq$ 10 GeV 

Set I corresponds to "no-tag" experiment when the forward and backward 
electrons at very small angles (less than 0.1 degree) in the dominant 
final state configuration are not detected.

\subsection{Total cross sections and distributions}

In the SM with $LL-$ contact term 110 tree level diagrams for the process 
$e^+ e^- \rightarrow e^+ e^- \mu^+ \mu^-$ can be generated. In order to 
optimize the procedure of calculation we separate them into subsets. Each 
subset contains subgraph corresponding to (with in-(out-) particles taken 
on-shell) some gauge invariant process of lower order. Detailed 
description of this procedure can be found in \cite{subsets}. 
We select the subset of two diagrams with t-channel photons (multiperipheral
diagrams) in the SM case and for the case of contact terms we add to them
eight diagrams with one t-channel photon and one contact interaction 
vertex (10 diagrams, see Fig.1).
The contributions from these subsets are generally speaking not 
always dominant in the
overall complete tree level set (under some conditions
single resonant diagrams with $Z$ boson in s-channel are not small), 
but usually about one order of magnitude larger than others.

The calculation of multiperipheral amplitudes containing t-channel 
photons is known to be very untrivial \cite{vermaseren}, especially in
the case when no cuts are imposed on final electrons ("no-tag"
experiment, total rate is finite because $m_e \neq 0$) and gauge
cancellations between diagrams are extremely strong. We  
used CompHEP 3.2 \cite{comphep} and tested the results by means of
EXCALIBUR \cite{excalibur}. 
In CompHEP numerical stability of the double poles 
$1/t^2$ cancellation to single ones is preserved by using quadruple 
precision and special algorithms of phase space generation
\cite{ilyin}.  

At the compositeness scale $\Lambda=$1 TeV and the energy $\sqrt{s}=$ 
200 GeV total cross sections 
in $pb$ for the process $e^+ e^- \rightarrow e^+ e^- \mu^+ \mu^-$ are
shown in Table 1.

\begin{table}[t]
\begin{center}
\begin{tabular}{|c|c|c|c|}
\hline
\multicolumn{4}{|c|}{$\sigma(e^+e^- \rightarrow e^+e^-\mu^+\mu^-)$, pb,
                   {\bf set I} (loose cuts) } \\ \hline
$M(\mu \mu)$ cut (GeV)          &    30    &    60    &    85     \\ \hline
SM                             &    4.165  &   0.527   &   0.135    \\
SM+$LL-$                       &    4.180   &   0.535   &   0.142    \\
deviation  in \%               & 0.4        & 1.5      &    4.9   \\ 
$N$, see (2)                   & 62500      &  4400    & 400        \\
\hline
\hline 
\multicolumn{4}{|c|}{$\sigma(e^+e^- \rightarrow e^+e^-\mu^+\mu^-)$, pb,
                   {\bf set II} (strong cuts) } \\ \hline
$M(\mu \mu)$ cut (GeV)         &    30    &    60    &    85     \\ \hline
SM                             & 1.4$*10^{-2}$  &  0.56$*10^{-2}$ &  
0.24$*10^{-2}$  \\
SM+$LL-$                       & 1.6$*10^{-2}$  &  0.66$*10^{-2}$ &  
0.29$*10^{-2}$  \\
deviation  in \%               & 14       &   18     &    21     \\ 
$N$, see (2)                   &    50    &   30     &    20     \\
\hline
\end{tabular}
\end{center} 
\caption{}
\end{table}

We used a very rough criteria (similar to criteria accepted in 
\cite{Schrempp}) that the number of events $N$ needed to observe 
the $\delta \sigma/\sigma$ fractional deviation from the SM cross section 
can be estimated by using the relation
\begin{equation}
\frac{\delta \sigma}{\sigma} \sim \frac{1}{\sqrt{N}} 
\end{equation}

i.e. $N$ is of order inverse fractional deviation squared. It follows from 
the Table that at the optimistic LEP2 integrated luminosity 500 $pb^{-1}$
the effect of $\Lambda=$ 1 TeV $LL-$ contact term cannot be observed in 
the total rate. For instance, 21\% effect in the case $M(\mu^+ \mu^-) \geq$
85 GeV, set II, requires the identification of 20 events while at LEP II
luminosity we have only one event per year. 

As usual in this sitiation, we inspected the influence of contact terms 
on the shape of various distributions, hoping that in the distributions
the effect could be much more pronounced if some phase space region is 
controlled strongly by contact interaction dynamics. We calculated 
and compared distributions over muon pair invariant mass
$d\sigma/dM_{\mu \bar \mu}$, muon angle 
$d\sigma/d\vartheta_{\mu}$, muon transverse momentum $d\sigma/dp_{t \, 
\mu}$, and muon energy $d\sigma/dE_{\mu}$. In this set of distributions 
for all cases the $LL-$ term effect looks like rather uniform background 
not changing significantly in the whole physical region of the process. 
For instance,
we show the distributions over muon angle in Fig.2. Their forward-
backward structure is of course completely
different from the central structure of $2 \rightarrow 2$ body reaction $e^+ 
e^- \rightarrow \mu^+ \mu^-$ \cite{Eichten,Schrempp} (where only partial 
wave with angular 
momentum zero contributes), but the shape of $2 \rightarrow 4$ body 
distribution $e^+ e^- \rightarrow e^+ e^- \mu^+ \mu^-$ with contact term 
is similar to standard distribution. 

\unitlength=0.60pt

\begin{figure}[h]
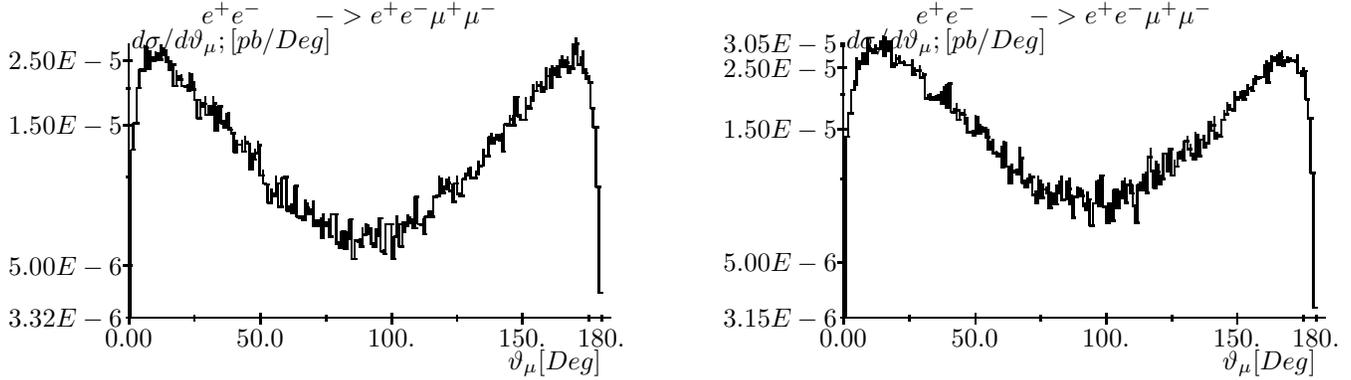

\begin{minipage}[h]{65mm}
\input{fig2.tex}
\end{minipage} 
\hfill
\begin{minipage}[h]{65mm}
\input{fig3.tex}
\end{minipage}
\caption{Left figure - $d\sigma/d\vartheta_{\mu}$, Standard Model, set II; 
right figure - $d\sigma/d\vartheta_{\mu}$, SM+$LL-$ contact term
 ($\Lambda=1$ TeV, set II) }
\end{figure}

We calculated also the fractional deviations $d\sigma_{LL-}/d\sigma_{SM}$
for the cases of loose (set I) and strong (set II) cuts.
The accuracy of our Monte Carlo (MC) calculation of the total rate is
around  
0.5\%. The accuracy of distributions is quite satisfactory for the most
important regions of the phase space (several percent in one bin). The
error in the ratio of distributions is of course more sensitive to these
statistical mistakes. Fig.3 shows that the accuracy of our MC is not
sufficient to show the 0.4\% effect in the ratio $(d\sigma_{LL-}/dM)/(
d\sigma_{SM}/dM)$ for the case of loose cuts. Of course for practical
purpose we do not need so precise calculation in so far as at LEP2 only
2000 events could be observed while 60000 are necessary (see Table 1). The
effect of contact term
could be clearly separated (Fig. 4) in the same ratio for the case of
strong cuts
(set II), but here we need the luminosity of order $10^4$ $pb^{-1}$ for
experimental observation.

\section{Distributions in the Standard Model with $LL-$ contact term
for the process $e^- p \rightarrow e^- \mu^+ \mu^- X$}

In the case of deep inelastic scattering we are using the MRS
parametrization of proton structure functions \cite{MRS},
developed on the basis of latest experimental data from HERA.
Available parametrizations of proton structure functions
can be used at the $Q^2$ scale sufficiently large, so
the calculations for the process $e^- q \rightarrow e^- \mu^+ \mu^- q$
were performed applying $|Q|= 3$ GeV cut for the momentum
transferred from the constituent quark.
Muon energy cut is equal to 10 GeV and we used 30 GeV for the muon pair
invariant mass cut. For HERA $ep$ collider the energy $\sqrt{s}=314$ GeV,
the electron-positron center of mass system is moving in the laboratory
system with the rapidity $y=1.654$
and the integrated luminosity at present time is several $pb^{-1}$.

\begin{table}[h]
\begin{center}
\begin{tabular}{|c|c|c|c|c|c|c|c|}
\hline
initial state&$eu$&$ed$&$e\bar u$&$e\bar d$&$es,e\bar s$&$ec,e\bar c$ &
total
\\ \hline
SM                     & 35.88& 3.24 & 1.16 & 0.52& 0.46  & 0.50& 41.76 \\
\hline
SM+$LL-$               & 36.19& 3.25 & 1.17 & 0.52& 0.47  & 0.50& 42.10
\\
\hline
\end{tabular}   
\end{center}  
\caption{Total cross sections ({\it fb}) for partonic spieces in the
process
$ e^- p \rightarrow e^- \mu^+ \mu^- X$, $q=u,d,s,c$,
$M(\mu^+ \mu^-) \geq 30$ GeV, $\Lambda=1$ TeV. }
\end{table}

\newpage

\unitlength=0.60pt

\begin{figure}[h]
\begin{minipage}[h]{70mm}
{\def\chepscale{0.75} 
\unitlength=\chepscale pt
\SetWidth{0.7}      
\SetScale{\chepscale}
\normalsize    
\begin{picture}(300,200)(0,0)
\Text(168.2,199.4)[t]{$e^+ e^-   -> e^+  e^-  \mu^+ \mu^-$}
\LinAxis(45.90,36.72)(290.82,36.72)(3.400,5,-4,-2.000,1.5)
\Text(74.7,29.9)[t]{$50$}
\Text(146.9,29.9)[t]{$100$}
\Text(218.6,29.9)[t]{$150$}
\Text(290.8,29.9)[t]{$200$}
\Text(290.8,20.3)[rt]{$ M_{\mu \mu}[GeV]$}
\LinAxis(45.90,36.72)(45.90,176.27)(4.000,5,4,-0.000,1.5)
\Text(39.2,36.7)[r]{$0.0$}
\Text(39.2,71.8)[r]{$0.5$}
\Text(39.2,106.2)[r]{$1$}
\Text(39.2,141.2)[r]{$1.5$}
\Text(39.2,176.3)[r]{$2.0$}
\rText(21.3,176.3)[tr][l]{$d\sigma_{LL-}/d\sigma_{SM}$}
\Line(283.3,129.9)(283.3,46.9) 
\Line(276.2,88.1)(290.8,88.1) 
\Line(269.1,173.4)(269.1,92.1) 
\Line(261.6,132.8)(276.2,132.8) 
\Line(254.5,165.5)(254.5,102.8) 
\Line(247.4,134.5)(261.6,134.5) 
\Line(240.3,95.5)(240.3,72.9) 
\Line(232.8,84.2)(247.4,84.2) 
\Line(225.7,119.8)(225.7,94.9) 
\Line(218.6,107.3)(232.8,107.3) 
\Line(211.5,140.1)(211.5,115.8) 
\Line(204.0,127.7)(218.6,127.7) 
\Line(196.9,119.8)(196.9,104.0) 
\Line(189.8,111.9)(204.0,111.9) 
\Line(182.8,122.0)(182.8,107.9) 
\Line(175.2,114.7)(189.8,114.7) 
\Line(168.2,125.4)(168.2,113.0) 
\Line(161.1,119.2)(175.2,119.2) 
\Line(153.5,122.6)(153.5,111.9) 
\Line(146.5,116.9)(161.1,116.9) 
\Line(139.4,110.7)(139.4,103.4) 
\Line(132.3,106.8)(146.5,106.8) 
\Line(124.8,109.0)(124.8,102.8) 
\Line(117.7,106.2)(132.3,106.2) 
\Line(110.6,105.6)(110.6,100.6) 
\Line(103.5,103.4)(117.7,103.4) 
\Line(96.0,108.5)(96.0,104.5) 
\Line(88.9,106.2)(103.5,106.2) 
\Line(81.8,111.3)(81.8,107.9) 
\Line(74.7,109.6)(88.9,109.6) 
\Line(67.2,108.5)(67.2,105.6) 
\Line(60.1,107.3)(74.7,107.3) 
\Line(53.0,107.9)(53.0,106.2) 
\Line(45.9,106.8)(60.1,106.8) 
\end{picture}
}
\end{minipage}
\hfill
\begin{minipage}[h]{70mm}
{\def\chepscale{0.75} 
\unitlength=\chepscale pt
\SetWidth{0.7}      
\SetScale{\chepscale}
\normalsize    
\begin{picture}(300,200)(0,0)
\Text(168.2,199.4)[t]{$e^+  e^-   -> e^+  e^-  \mu^+ \mu^-$}
\LinAxis(45.90,36.72)(290.82,36.72)(3.200,5,-4,1.000,1.5)
\Text(107.2,29.9)[t]{$50$}
\Text(183.6,29.9)[t]{$100$}
\Text(260.4,29.9)[t]{$150$}
\Text(290.8,20.3)[rt]{$\vartheta_{\mu} [Deg]$}
\LinAxis(45.90,36.72)(45.90,176.27)(4.000,10,4,0.000,1.5)
\Text(39.2,36.7)[r]{$0.80$}
\Text(39.2,71.8)[r]{$0.90$}
\Text(39.2,106.8)[r]{$1.00$}
\Text(39.2,141.2)[r]{$1.10$}
\Text(39.2,176.3)[r]{$1.20$}
\rText(16.3,176.3)[tr][l]{$d\sigma_{LL-}/d\sigma_{SM}$}
\Line(282.9,130.5)(282.9,104.5) 
\Line(275.4,117.5)(290.8,117.5) 
\Line(267.5,137.3)(267.5,115.3) 
\Line(259.9,126.6)(275.4,126.6) 
\Line(252.4,106.2)(252.4,85.9) 
\Line(244.5,96.0)(259.9,96.0) 
\Line(237.0,143.5)(237.0,119.8) 
\Line(229.5,131.6)(244.5,131.6) 
\Line(221.6,112.4)(221.6,89.3) 
\Line(214.0,100.6)(229.5,100.6) 
\Line(206.5,115.3)(206.5,92.1) 
\Line(198.6,103.4)(214.0,103.4) 
\Line(191.1,89.8)(191.1,67.2) 
\Line(183.6,78.5)(198.6,78.5) 
\Line(175.7,126.6)(175.7,101.7) 
\Line(168.2,114.1)(183.6,114.1) 
\Line(160.6,114.7)(160.6,84.2) 
\Line(152.7,99.4)(168.2,99.4) 
\Line(145.2,137.9)(145.2,110.7) 
\Line(137.7,124.3)(152.7,124.3) 
\Line(129.8,118.6)(129.8,93.8) 
\Line(122.3,106.2)(137.7,106.2) 
\Line(114.7,137.3)(114.7,112.4) 
\Line(106.8,124.9)(122.3,124.9) 
\Line(99.3,109.6)(99.3,87.6) 
\Line(91.8,98.3)(106.8,98.3) 
\Line(83.9,131.1)(83.9,108.5) 
\Line(76.4,119.8)(91.8,119.8) 
\Line(68.8,119.8)(68.8,101.1) 
\Line(60.9,110.7)(76.4,110.7) 
\Line(53.4,103.4)(53.4,83.6) 
\Line(45.9,93.2)(60.9,93.2) 
\end{picture}
}
\end{minipage}
\caption{Left figure - 
ratio $d\sigma_{LL-}/d\sigma_{SM}$ for the muon pair invariant
mass, $\Lambda=1$ TeV, set I; right figure - 
ratio $d\sigma_{LL-}/d\sigma_{SM}$ for the muon pair angle,
$\Lambda=1$ TeV, set I. The error of Monte Carlo calculation 
in the ratio is indicated. }
\end{figure}
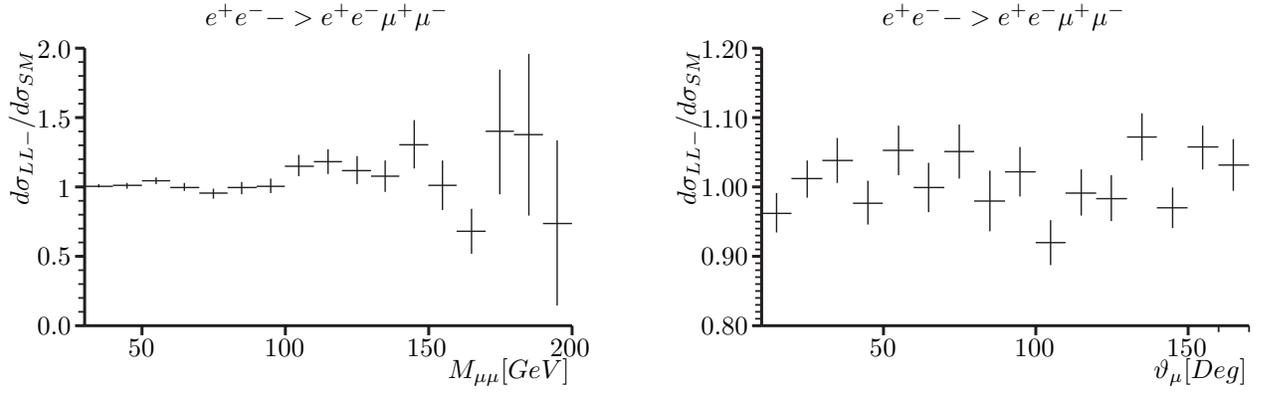

\begin{figure}[h]
\begin{minipage}[h]{70mm}
{\def\chepscale{0.75} 
\unitlength=\chepscale pt
\SetWidth{0.7}      
\SetScale{\chepscale}
\normalsize    
\begin{picture}(300,200)(0,0)
\Text(168.2,199.4)[t]{$e^+  e^-   -> e^+  e^-  \mu^+ \mu^-$}
\LinAxis(45.90,36.72)(290.82,36.72)(3.333,3,-4,-1.000,1.5)
\Text(70.5,29.9)[t]{$90$}
\Text(143.9,29.9)[t]{$120$}
\Text(217.4,29.9)[t]{$150$}
\Text(290.8,29.9)[t]{$180$}
\Text(290.8,20.3)[rt]{$M_{\mu \mu} [GeV]$}
\LinAxis(45.90,36.72)(45.90,176.27)(4.000,2,4,-0.000,1.5)
\Text(39.2,36.7)[r]{$0.8$}
\Text(39.2,71.8)[r]{$1.0$}
\Text(39.2,106.8)[r]{$1.2$}
\Text(39.2,141.2)[r]{$1.4$}
\Text(39.2,176.3)[r]{$1.6$}
\rText(21.3,176.3)[tr][l]{$d\sigma_{LL-}/d\sigma_{SM}$}
\Line(278.3,176.8)(278.3,123.7) 
\Line(266.2,161.6)(290.8,161.6) 
\Line(253.7,114.7)(253.7,88.7) 
\Line(241.6,101.7)(266.2,101.7) 
\Line(229.5,116.9)(229.5,96.6) 
\Line(217.0,106.8)(241.6,106.8) 
\Line(204.9,116.9)(204.9,102.8) 
\Line(192.8,109.6)(217.0,109.6) 
\Line(180.3,102.8)(180.3,93.2) 
\Line(168.2,97.7)(192.8,97.7) 
\Line(156.1,106.2)(156.1,96.0) 
\Line(143.5,101.1)(168.2,101.1) 
\Line(131.4,116.4)(131.4,107.3) 
\Line(119.3,111.9)(143.5,111.9) 
\Line(106.8,113.6)(106.8,106.8) 
\Line(94.7,110.2)(119.3,110.2) 
\Line(82.6,111.3)(82.6,105.6) 
\Line(70.1,108.5)(94.7,108.5) 
\Line(58.0,106.8)(58.0,100.0) 
\Line(45.9,103.4)(70.1,103.4) 
\end{picture}
}
\end{minipage}
\hfill
\begin{minipage}[h]{70mm}
{\def\chepscale{0.75} 
\unitlength=\chepscale pt
\SetWidth{0.7}      
\SetScale{\chepscale}
\scriptsize    
\begin{picture}(300,200)(0,0)
\Text(164.8,199.4)[t]{$e^+  e^-   -> e^+  e^-  \mu^+ \mu^-$}
\LinAxis(37.55,30.17)(292.49,30.17)(3.600,5,-4,-0.000,1.5)
\Text(37.6,24.6)[t]{$0.0$}
\Text(108.5,24.6)[t]{$50$}
\Text(179.0,24.6)[t]{$100$}
\Text(249.9,24.6)[t]{$150$}
\Text(292.5,16.8)[rt]{$\vartheta_{\mu} [Deg]$}
\LinAxis(37.55,30.17)(37.55,180.45)(3.000,5,4,0.000,1.5)
\Text(31.7,30.2)[r]{$0.5$}
\Text(31.7,80.4)[r]{$1$}
\Text(31.7,130.2)[r]{$1.5$}
\Text(31.7,180.4)[r]{$2.0$}
\rText(17.1,180.4)[tr][l]{$d\sigma_{LL-}/d\sigma_{SM}$}
\Line(285.4,91.1)(285.4,83.8) 
\Line(278.3,87.2)(292.5,87.2) 
\Line(271.2,93.3)(271.2,87.7) 
\Line(264.1,90.5)(278.3,90.5) 
\Line(257.0,94.4)(257.0,88.8) 
\Line(249.9,91.6)(264.1,91.6) 
\Line(242.8,92.2)(242.8,86.6) 
\Line(235.7,89.4)(249.9,89.4) 
\Line(228.7,97.2)(228.7,89.9) 
\Line(221.6,93.3)(235.7,93.3) 
\Line(214.5,105.0)(214.5,96.6) 
\Line(207.4,101.1)(221.6,101.1) 
\Line(200.3,106.7)(200.3,97.8) 
\Line(193.2,102.2)(207.4,102.2) 
\Line(186.1,110.1)(186.1,99.4) 
\Line(179.0,104.5)(193.2,104.5) 
\Line(171.9,116.2)(171.9,106.7) 
\Line(164.8,111.2)(179.0,111.2) 
\Line(157.7,126.3)(157.7,115.1) 
\Line(150.6,120.7)(164.8,120.7) 
\Line(143.5,116.8)(143.5,107.3) 
\Line(136.4,111.7)(150.6,111.7) 
\Line(129.3,123.5)(129.3,114.0) 
\Line(122.3,118.4)(136.4,118.4) 
\Line(115.2,134.1)(115.2,124.6) 
\Line(108.1,129.6)(122.3,129.6) 
\Line(101.0,120.7)(101.0,111.2) 
\Line(93.9,115.6)(108.1,115.6) 
\Line(86.8,109.5)(86.8,102.8) 
\Line(79.7,106.1)(93.9,106.1) 
\Line(72.6,118.4)(72.6,111.2) 
\Line(65.5,115.1)(79.7,115.1) 
\Line(58.4,101.1)(58.4,96.1) 
\Line(51.3,98.9)(65.5,98.9) 
\Line(44.2,91.1)(44.2,84.4) 
\Line(37.6,87.7)(51.3,87.7) 
\end{picture}
}
\end{minipage}
\caption{ Left figure -
ratio $d\sigma_{LL-}/d\sigma_{SM}$ for the muon pair invariant
mass, $\Lambda=1$ TeV, set II, right figure - 
ratio $d\sigma_{LL-}/d\sigma_{SM}$ for the muon angle,
 $\Lambda=1$ TeV, set II. The error of Monte Carlo calculation in the 
ratio is indicated. } 
\end{figure}
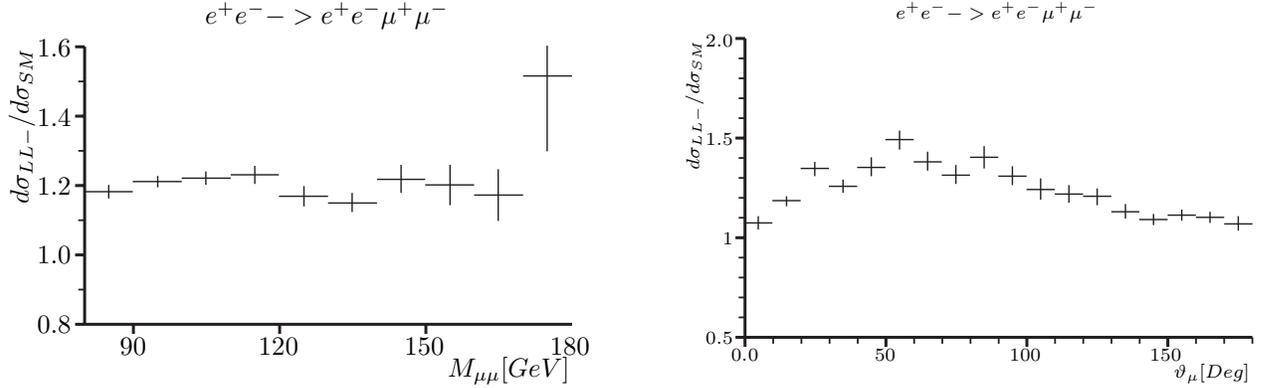

Total cross sections for valence and sea quarks are shown in Table 2.
Similar to $e^+ e^-$ case with loose cuts (set I), the contribution of
the contact term in $ep$ scattering is very small. According
to criteria (2) in order to observe the deviation in total rate of order 
1\% it is necessary to identify approximately $10^4$ events, while even
at upgraded high luminosity HERA ($L \sim 10^2 \, pb^{-1}$) it would be
possible to observe of order $10^1$ events. We show the 
fractional deviations of the muon pair invariant mass distribution and 
muon angle distribution in Fig.5. In the distributions the
effect is also practically unobservable. 

\section{Conclusion}

We calculated the effect of $LL-$ contact term in the four-fermion channel
$e^+ e^- \rightarrow e^+ e^- \mu^+ \mu^-$ at the energy 200 GeV.
Search strategies with loose and strong cuts imposed on the final state
were considered.

\newpage

\begin{figure}[h]
\begin{minipage}[h]{70mm}
{\def\chepscale{0.75} 
\unitlength=\chepscale pt
\SetWidth{0.7}      
\SetScale{\chepscale}
\normalsize    
\begin{picture}(300,200)(0,0)
\Text(168.2,199.4)[t]{$e^-  p    -> e^-  \mu^+  \mu^-  X$}
\LinAxis(45.90,36.72)(290.82,36.72)(3.333,3,-4,0.000,1.5)
\Text(45.9,29.9)[t]{$30$}
\Text(119.3,29.9)[t]{$60$}
\Text(192.8,29.9)[t]{$90$}
\Text(266.2,29.9)[t]{$120$}
\Text(290.8,20.3)[rt]{$M_{\mu \mu}[Gev]$}
\LinAxis(45.90,36.72)(45.90,176.27)(4.062,2,4,1.000,1.5)
\Text(39.2,53.7)[r]{$0.8$}
\Text(39.2,88.1)[r]{$1.0$}
\Text(39.2,122.6)[r]{$1.2$}
\Text(39.2,157.1)[r]{$1.4$}
\rText(21.3,176.3)[tr][l]{$d\sigma_{LL-}/s\sigma_{SM}$}
\Line(278.3,169.5)(278.3,36.7) 
\Line(266.2,88.7)(290.8,88.7) 
\Line(253.7,176.3)(253.7,69.5) 
\Line(241.6,123.2)(266.2,123.2) 
\Line(229.5,136.2)(229.5,79.1) 
\Line(217.0,107.3)(241.6,107.3) 
\Line(204.9,73.4)(204.9,49.7) 
\Line(192.8,61.6)(217.0,61.6) 
\Line(180.3,113.6)(180.3,84.2) 
\Line(168.2,98.9)(192.8,98.9) 
\Line(156.1,103.4)(156.1,84.2) 
\Line(143.5,93.8)(168.2,93.8) 
\Line(131.4,110.7)(131.4,93.8) 
\Line(119.3,102.3)(143.5,102.3) 
\Line(106.8,93.2)(106.8,82.5) 
\Line(94.7,88.1)(119.3,88.1) 
\Line(82.6,89.3)(82.6,82.5) 
\Line(70.1,85.9)(94.7,85.9) 
\Line(58.0,91.0)(58.0,87.0) 
\Line(45.9,88.7)(70.1,88.7) 
\end{picture}
}
\end{minipage}
\hfill
\begin{minipage}[h]{70mm}   
{\def\chepscale{0.75} 
\unitlength=\chepscale pt
\SetWidth{0.7}      
\SetScale{\chepscale}
\normalsize    
\begin{picture}(300,200)(0,0)
\Text(168.2,199.4)[t]{$e^-  p    -> e^-  \mu^+  \mu^-  X$}
\LinAxis(45.90,36.72)(290.82,36.72)(4.000,3,-4,1.000,1.5)
\Text(86.8,29.9)[t]{$30$}
\Text(148.1,29.9)[t]{$60$}
\Text(209.0,29.9)[t]{$90$}
\Text(270.4,29.9)[t]{$120$}
\Text(290.8,20.3)[rt]{$\vartheta_{\mu} [Deg]$}
\LinAxis(45.90,36.72)(45.90,176.27)(4.000,10,4,0.000,1.5)
\Text(39.2,36.7)[r]{$0.80$}
\Text(39.2,71.8)[r]{$0.90$}
\Text(39.2,106.8)[r]{$1.00$}
\Text(39.2,141.2)[r]{$1.10$}
\Text(39.2,176.3)[r]{$1.20$}
\rText(16.3,176.3)[tr][l]{$d\sigma_{LL-}/d\sigma_{SM}$}
\Line(280.4,124.9)(280.4,36.7) 
\Line(270.4,74.6)(290.8,74.6) 
\Line(259.9,132.2)(259.9,65.5) 
\Line(249.9,98.9)(270.4,98.9) 
\Line(239.5,142.9)(239.5,84.2) 
\Line(229.5,113.6)(249.9,113.6) 
\Line(219.1,143.5)(219.1,91.0) 
\Line(209.0,116.9)(229.5,116.9) 
\Line(198.6,137.9)(198.6,92.1) 
\Line(188.6,114.7)(209.0,114.7) 
\Line(178.2,116.9)(178.2,84.7) 
\Line(168.2,101.1)(188.6,101.1) 
\Line(158.1,109.0)(158.1,78.5) 
\Line(147.7,93.8)(168.2,93.8) 
\Line(137.7,114.7)(137.7,89.3) 
\Line(127.3,101.7)(147.7,101.7) 
\Line(117.2,127.1)(117.2,106.2) 
\Line(106.8,116.4)(127.3,116.4) 
\Line(96.8,107.3)(96.8,92.1) 
\Line(86.4,99.4)(106.8,99.4) 
\Line(76.4,127.7)(76.4,114.7) 
\Line(65.9,120.9)(86.4,120.9) 
\Line(55.9,111.9)(55.9,101.1) 
\Line(45.9,106.2)(65.9,106.2) 
\end{picture}
}
\end{minipage}
\caption{Left figure -
ratio $d\sigma_{LL-}/d\sigma_{SM}$ for the muon pair
invariant mass, $\Lambda=1$ TeV; right figure -
ratio $d\sigma_{LL-}/d\sigma_{SM}$ for the muon angle,
$\Lambda=1$ TeV.}
\end{figure}
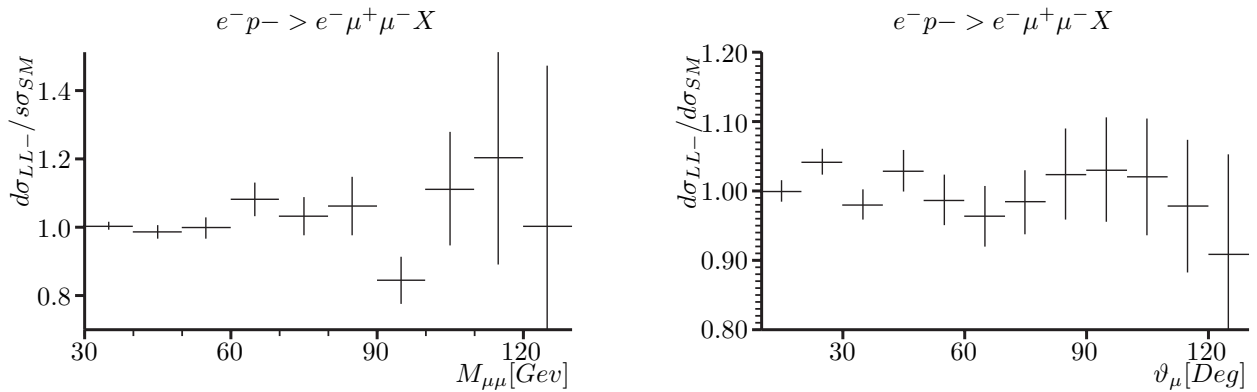

In the case of loose cuts (set I)
at the compositeness scale 1 TeV
the difference in the total rates is around 1\%. It would be hardly
possible to observe the deviations from the SM in the distibutions.
In the case of strong cuts (set II) the effect of contact terms is much
more
pronounced and is of order 20\% in the total rate and could be clearly
observed in the distributions,
but the number of events at LEP2 luminosity of several hundred $pb^{-1}$
is
too small. Separation of the contact
term is possible at the integrated luminosity of order 10 $fb^{-1}$.

The deviation from the SM distributions caused by contact terms is 
rather uniform and in all cases considered it looks like some 
bias of constant level in the whole physical region.
At the compositeness scale 4 TeV the difference of SM and SM+$LL-$
distributions in the same four fermionic channel decreases approximately by 
one order of magnitude.  

\begin{table}[h]
\begin{center}
\begin{tabular}{|l|c|c|c|} \hline
            & $\sigma_{tot}$ (pb) & deviation in & deviation in \\
                &      SM          & $\sigma_{tot}$ & $d\sigma
                                                    /dcos\vartheta_{\mu}$ \\
\hline
 $e^+ e^- \rightarrow \mu^+ \mu^-$     & 3.0 & about 300\% & up to 300\%  \\ 
$e^+ e^- \rightarrow e^+ e^- \mu^+ \mu^-$, set I & 4.2 & 0.4\%& negligible  \\
$e^+ e^- \rightarrow e^+ e^- \mu^+ \mu^-$, set II& 2*$10^{-2}$& 15\% & up 
                                                              to 50\% \\ \hline
\end{tabular}
\end{center}
\caption{Typical deviations in the total rate and muon angular distribution
          of the reactions $e^+ e^- \rightarrow \mu^+ \mu^-$ and 
            $e^+ e^- \rightarrow e^+ e^- \mu^+ \mu^-$ caused by $LL-$ 
           contact term at the energy ${\protect \sqrt{s}=200}$ GeV and 
            compositeness scale $\Lambda=$ 1 TeV. Muon pair mass $M_{\mu 
            \mu} \geq$ 30 GeV in the case of set I,II.} 
\end{table}

We calculated also the "four-fermion" channel $e^- q  \rightarrow e^-
\mu^+ \mu^- q$ at the energy of HERA $\sqrt{s}=314$ GeV. The effect
of the contact term in the total rate at the compositeness scale 1 TeV is 
about 1\%. Again, it would be hardly possible to observe any deviations
from the SM distributions.

Four-fermion channel considered by us does not show new critical
advantages over the possibilities of the compositeness search 
considered earlier \cite{ee}. We compare the magnitude of the effect for the
reactions $e^+ e^- \rightarrow \mu^+ \mu^-$ and $e^+ e^- \rightarrow 
e^+ e^- \mu^+ \mu^-$ in Table 3. The discovery potential of four-fermion 
reactions is critically dependent from the collider luminosity. 

\begin{center}
{\bf Acknowledgement}
\end{center}
The research of M.D. was partially supported by RFBR 96-02-19773a, 
St-Pb. grant 95-0-6.4-38 and INTAS 93-1180ext.

\end{document}